\documentclass{llncs}

\usepackage{epsfig}

\newcommand{\comment}[1]{}
\newcommand{\Order}{\mathcal{O}}
\newcommand{\order}{o}

\title{CRAM: Compressed Random Access Memory}

\author{
  Jesper Jansson\inst{1}
\and
  Kunihiko Sadakane\inst{2}
\and
  Wing-Kin Sung\inst{3}
}

\institute{
  Ochanomizu University, 2-1-1 Otsuka, Bunkyo-ku, Tokyo 112-8610, Japan. \\
  ~E-mail: \texttt{jesper.jansson@ocha.ac.jp}
\and
  National Institute of Informatics, 2-1-2 Hitotsubashi, Chiyoda-ku,
  Tokyo 101-8430, Japan.
  ~E-mail: \texttt{sada@nii.ac.jp}
\and
  National University of Singapore, 13~Computing Drive, Singapore~117417. \\
  ~E-mail: \texttt{ksung@comp.nus.edu.sg}
}

\date{}

\begin{document}

\maketitle

\begin{abstract}
We present a new data structure called the
\emph{Compressed Random Access Memory}~(CRAM)
that can store a dynamic string~$T$ of characters, e.g., representing the
memory of a computer, in compressed form while achieving asymptotically
almost-optimal bounds (in terms of empirical entropy) on the compression
ratio.
It allows short substrings of~$T$ to be decompressed and retrieved efficiently
and, significantly, characters at arbitrary positions of~$T$ to be modified
quickly during execution \emph{without decompressing the entire string}.
This can be regarded as a new type of data compression that can update
a compressed file directly.
Moreover, at the cost of slightly increasing the time spent per operation,
the CRAM
can be extended to also support insertions and deletions.
Our key observation that the empirical entropy of a string does not change
much after a small change to the string, as well as our simple yet efficient
method for maintaining an array of variable-length blocks under length
modifications, may be useful for many other applications as well.
\end{abstract}


\section{Introduction}

Certain modern-day information technology-based applications require
random access to very large data structures.
For example, to do genome assembly in bioinformatics, one needs to maintain
a huge graph~\cite{SimpsonWongJackmanScheinJonesBirol2009}.
Other examples include dynamic programming-based problems, such as optimal
sequence alignment or finding maximum bipartite matchings, which need to
create large tables (often containing a lot of redundancy).
Yet another example is in image processing, where one sometimes needs to
edit a high-resolution image which is too big to load into the main memory
of a computer all at once.
Additionally, a current trend in the mass consumer electronics market is
cheap mobile devices with limited processing power and relatively small
memories; although these are not designed to process massive amounts of data,
it could be economical to store non-permanent data and software on them more
compactly,~if~possible.

The standard solution to the above problem is to employ secondary memory
(disk storage, etc.) as an extension of the main memory of a computer.
This technique is called \emph{virtual memory}.
The drawback of virtual memory is that the processing time will be slowed
down since accessing the secondary memory is an order of magnitude slower
than accessing the main memory.
An alternative approach is to compress the data~$T$ and store it in the main
memory.
By using existing data compression methods, $T$~can be stored in
$n H_k + \order(n \log \sigma)$-bits space~\cite{FerMan05,GGV03}
for every $0 \le k < \log_\sigma n$,
where $n$ is the length of~$T$, $\sigma$ is the size of the alphabet,
and $H_k(T)$ denotes the $k$-th order empirical entropy of~$T$.
Although greatly reducing the amount of storage needed, it does not work well
because it becomes computationally expensive to access and update~$T$.

Motivated by applications that would benefit from having a large virtual
memory that supports fast access- and update-operations, we consider
the following task:
Given a memory/text $T[1..n]$ over an alphabet of size $\sigma$, maintain
a data structure that stores $T$ compactly while supporting the following
operations.
(We assume that $\ell = \Theta(\log_\sigma n)$ is the length of one machine
word.)
\begin{itemize}
\vspace*{-1mm}
\item[{\raise0.3pt\hbox{$\bullet$}}]
  \textnormal{\texttt{access}}$(T,i)$:
  Return the substring $T[i..(i+\ell-1)]$.
\item[{\raise0.3pt\hbox{$\bullet$}}]
  \textnormal{\texttt{replace}}$(T,i,c)$:
  Replace $T[i]$ by a character $c \in [\sigma]$.
  \footnote{The notation $[\sigma]$ stands for the
  set $\{1,2,\ldots,\sigma\}$.}
\item[{\raise0.3pt\hbox{$\bullet$}}]
  \textnormal{\texttt{delete}}$(T,i)$:
  Delete $T[i]$,
  i.e., make $T$ one character shorter.
\item[{\raise0.3pt\hbox{$\bullet$}}]
  \textnormal{\texttt{insert}}$(T,i,c)$:
  Insert a character $c$ into $T$ between
  positions $i-1$ and $i$,
  i.e., make $T$ one character longer.
\end{itemize}

\noindent
\textbf{Compressed Read Only Memory:}
When only the \texttt{access} operation is supported,
we call the data structure \emph{Compressed Read Only Memory}.
Sadakane and Grossi \cite{SG_SODA2006},
Gonz\'alez and Navarro \cite{GonzalezN06}, and
Ferragina and Venturini \cite{FerVen07b}
developed storage schemes for storing a text succinctly that
allow constant-time access to any word in the text.
More precisely, these schemes store $T[1..n]$ in
$nH_k + \Order\left(n \log \sigma \left(\frac{k \log \sigma + \log \log n}{\log n}\right)\right)$ bits\footnote{Reference~\cite{SG_SODA2006} has
a slightly worse space complexity.}
and \texttt{access}$(T,i)$ takes $\Order(1)$ time,
and both the space and access time are optimal for this task.
Note, however, that none of these schemes allow $T$ to be modified.

\medskip

\noindent
\textbf{Compressed Random Access Memory (CRAM):}
When the operations \texttt{access} and \texttt{replace} are supported,
we call the data structure \emph{Compressed Random Access Memory} (CRAM).
As far as we know, it has not been considered previously in the literature,
even though it appears to be a fundamental and important data structure.

\medskip

\noindent
\textbf{Extended CRAM:}
When all four operations are supported,
we call the data structure \emph{extended CRAM}.
It is equivalent to \emph{the dynamic array}~\cite{RRR_WADS2001}
and also solves
\emph{the list representation problem}~\cite{FredmanS_STOC1989}.
Fredman and Saks \cite{FredmanS_STOC1989} proved
a cell probe lower bound of $\Omega(\log n/ \log \log n)$ time 
for the latter, and also showed that $n^{\Omega(1)}$ update time is needed
to support constant-time \texttt{access}.
Raman \emph{et al.}~\cite{RRR_WADS2001}
presented an $n \log \sigma + \order(n \log \sigma)$-bit data structure
which supports \texttt{access}, \texttt{replace}, \texttt{delete}, and
\texttt{insert} in $\Order(\log n/\log \log n)$ time.
Navarro and Sadakane \cite{NavSad10} recently gave a data structure using
$nH_0(T) + \Order(n\log\sigma/\log^\epsilon n + \sigma \log^\epsilon n)$
bits that supports \texttt{access}, \texttt{delete}, and \texttt{insert}
in $\Order(\frac{\log n}{\log\log n}(1+\frac{\log\sigma}{\log\log n}))$ time.

\subsection{Our contributions}

This paper studies the complexity of maintaining the CRAM and
extended CRAM data structures.
We assume the uniform-cost word RAM model with~word size
$w = \Theta(\log n)$ bits,
i.e., 
standard arithmetic and bitwise boolean operations
on $w$-bit word-sized operands can be performed in constant time~\cite{Hag98}.
Also, we assume that the memory consists of a sequence of
bits, and each bit is identified with an address in $0, \ldots, 2^w-1$.
Furthermore, any consecutive $w$ bits can be accessed in constant time.
(Note that this memory model is equivalent under the word RAM model to
a standard memory model consisting of a sequence of words of some fixed
length.)
At any time, if the highest address of the memory used by the algorithm
is~$s$, the space used by the algorithm is said to be
$s+1$~bits~\cite{HagerupR02}.

\smallskip

Our main results for the CRAM are summarized in:
\begin{theorem}\label{th:main1}
Given a text $T[1..n]$ over an alphabet of size~$\sigma$
and any fixed $\epsilon > 0$,
after $\Order(n\log\sigma/\log n)$ time preprocessing,
the CRAM data structure for~$T[1..n]$ can be stored in
$nH_k(T)+\Order\left(n\log\sigma \left((k+1)\epsilon + \frac{k\log\sigma+\log\log n}{\log n}\right)\right)$ bits
for every $0 \le k < \log_\sigma n$ simultaneously,
where $H_k(T)$ denotes the $k$-th order empirical entropy of~$T$,
while supporting \textnormal{\texttt{access}}$(T,i)$ in $\Order(1)$ time
and \textnormal{\texttt{replace}}$(T,i,c)$ for any character~$c$
in $\Order(1/\epsilon)$ time.
\end{theorem}

Theorem~\ref{th:main1} is proved in Section~\ref{sec:replace} below.

Next, by setting
$\epsilon = \max\{\frac{\log \sigma}{\log n}, \frac{\log\log n}{(k+1)\log n} \}$,
we obtain:
\begin{corollary}
Given a text $T[1..n]$ over an alphabet of size~$\sigma$
and any fixed $k = \order(\log_\sigma n)$,
after $\Order(n\log\sigma/\log n)$ time preprocessing,
the CRAM data structure for~$T[1..n]$ can be stored in
$nH_k(T)+\Order\left(n\log\sigma \cdot \frac{k\log\sigma+\log\log n}{\log n}\right)$ bits
while supporting \textnormal{\texttt{access}}$(T,i)$ in $\Order(1)$ time and
\textnormal{\texttt{replace}}$(T,i,c)$ for any character~$c$
in $\Order(\min\{\log_\sigma n, (k+1)\log n/\log\log n\})$ time.
\end{corollary}

\smallskip

For the extended CRAM, we have:
\begin{theorem}\label{th:main2}
Given a text $T[1..n]$ over an alphabet of size~$\sigma$,
after $\Order(n\log\sigma/\log n)$ time preprocessing,
the extended CRAM data structure for~$T[1..n]$ can be stored in
$nH_k(T)+\Order\left(n\log\sigma \cdot \frac{k\log\sigma+(k+1)\log\log n}{\log n}\right)$ bits
for every $0 \le k < \log_\sigma n$ simultaneously,
where $H_k(T)$ denotes the $k$-th order empirical entropy of~$T$,
while supporting all four operations in $\Order(\log n/\log\log n)$ time.
\end{theorem}

Due to lack of space, the proof of
Theorem~\ref{th:main2} is given in Appendix~\ref{sec:indel}.

\medskip

Table~\ref{tbl:comparison} shows a comparison with existing data structures.
Many existing dynamic data structures for storing compressed
strings~\cite{GNtcs09,HeMun10,MNtalg08,NavSad10}
use the fact $nH_0(S) = \log {{n}\choose{n_1,\ldots,n_\sigma}}$ where $n_c$
is the number of occurrences of character $c$ in the string $S$.
However, this approach is helpful for small alphabets only because of the size
of the auxiliary data.
For large alphabets, generalized wavelet trees~\cite{FMMNtalg07} can be used
to decompose a large alphabet into smaller ones, but this slows down the
access and update times.
For example, if $\sigma = \sqrt{n}$, the time complexity of those data
structures is $\Order((\log n/\log\log n)^2)$, 
while ours is $\Order(\log n/\log\log n)$, or even constant.
Also, a technical issue when using large alphabets is how to update the code
tables for encoding characters to achieve the entropy bound.
Code tables that achieve the entropy bound will change when the string
changes, and updating the entire data structure with the new code table is
too time-consuming.

\begin{table}[tb]
  \small
  \caption{Comparison of existing data structures and the new ones from this
  paper.
  For simplicity, we assume $\sigma = \order(n)$.
  The upper table lists results for the Compressed Read Only Memory (the
  first line) and the CRAM (the second and third lines), and the lower table
  lists results for the extended CRAM.}
  \label{tbl:comparison}

\smallskip
\footnotesize

  \begin{tabular}{ccll}
  \texttt{access} & \texttt{replace} & Space (bits) & Ref. \\ \hline
  $\Order(1)$ & --- & $nH_k(T)+\Order\left(n\log\sigma \cdot \frac{k\log\sigma+\log\log n}{\log n}\right)$ & \cite{FerVen07b,GonzalezN06} \\
  $\Order(1)$ & $\Order(\min\{\log_\sigma n, \frac{(k+1)\log n}{\log\log n}\})$ & $nH_k(T)+\Order\left(n\log\sigma \cdot \frac{k\log\sigma+\log\log n}{\log n}\right)$ & New \\
  $\Order(1)$ & $\Order(\frac{1}{\epsilon})$ & $nH_k(T)+\Order\left(n\log\sigma \left(\frac{k\log\sigma+\log\log n}{\log n} + (k+1)\epsilon \right)\right)$ & New \\ \hline
  \end{tabular}
\bigskip

  \begin{tabular}{cll}
  \texttt{access}/\texttt{replace}/\texttt{insert}/\texttt{delete}\,\,\, & Space (bits) & Ref. \\ \hline
  {$\Order(\frac{\log^2 n}{\log\sigma})$} & $nH_k(T) + \order(n \log \sigma)$ & \cite{NavSad10} \\
  {$\Order(\frac{\log\sigma\log n}{(\log\log n)^2})$} & $nH_0(T) + \Order\left(n \log \sigma \cdot \frac{1}{\log^\epsilon n}\right)$ & \cite{NavSad10} \\
  {$\Order(\frac{\log n}{\log\log n})$} & $nH_0(T)+\Order\left(n\log\sigma \cdot \frac{\log\log n}{\log n}\right)$ & New \\
  $\Order(\frac{\log n}{\log\log n})$ & $nH_k(T)+\Order\left(n\log\sigma \cdot \frac{k\log\sigma+(k+1)\log\log n}{\log n}\right)$ & New \\
  \hline
  \end{tabular}
\end{table}

Our results depend on a new analysis of the empirical entropies of
\emph{similar} strings in Section~\ref{sec:entropysimilar}.
We prove that \emph{the empirical entropy of a string does not change a lot
after a small change to the string} (Theorem~\ref{th:entropysimilar}).
By using this fact, we can delay updating the entire code table.
Thus, after each update operation to the string, we just change a part of
the data structure according to the new code table.
In Section~\ref{sec:replace}, we show that the redundancy in space usage by
this method is negligible, and we obtain Theorem~\ref{th:main1}.

Looking at Table~\ref{tbl:comparison}, we observe that Theorem~\ref{th:main1}
can be interpreted as saying that for arbitrarily small, fixed
$\epsilon > 0$, by spending $\Order(n \log \sigma \cdot \epsilon(k+1))$ bits
space more than the best existing data structures for
Compressed Read Only Memory, we can also get
$\Order(1/\epsilon)$ (i.e., constant) time \texttt{replace} operations.

\subsection{Organization of the paper}

Section~\ref{sec:preliminaries} reviews the definition of the empirical
entropy of a string and the data structure of Ferragina and
Venturini~\cite{FerVen07b}.
In Section~\ref{sec:entropysimilar}, we prove an important result on
the empirical entropies of similar strings.
In Section~\ref{sec:memory} and Appendix~\ref{sec:proof1}, we describe
a technique for maintaining an array of variable-length blocks.
Section~\ref{sec:replace} and Appendix~\ref{sec:indel} explain how to
implement the CRAM and the extended CRAM data structures to achieve
the bounds stated in Theorems~\ref{th:main1} and~\ref{th:main2} above.
Finally, Section~\ref{sec:conclusion} gives some concluding remarks.
Experimental results that demonstrate the good performance of the CRAM
in practice can be found in Appendix~\ref{sec:experiments}.


\section{Preliminaries}\label{sec:preliminaries}

\subsection{Empirical entropy}\label{sec:entropydef}

The compression ratio of a data compression method is often expressed in
terms of the \emph{empirical entropy} of the input strings~\cite{KosMan99}.
We first recall the definition of this concept.
Let $T$ be a string of length $n$ over an alphabet
${\cal A} = [\sigma]$.
Let $n_c$ be the number of occurrences of $c \in {\cal A}$ in $T$.  
Let $\{P_c = n_c / n\}_{c=1}^\sigma$ be the empirical probability distribution for the string $T$.
The $0$-th order empirical entropy of $T$ is defined as
$H_0(T) = -\sum_{c=1}^\sigma P_c \log P_c.$
We also use $H_0(p)$ to denote the $0$-th order empirical entropy of a string
whose empirical probability distribution is $p$.

Next, let $k$ be any non-negative integer.
If a string $s \in {\cal A}^k$ precedes a symbol $c$ in $T$,
$s$ is called the \emph{context of $c$}.
We denote by $T^{(s)}$ the string that is the concatenation of 
all symbols, each of whose context in $T$ is $s$.
The $k$-th order empirical entropy of $T$ is defined as
$H_k(T) = \frac{1}{n} \sum_{s \in {\cal A}^k} |T^{(s)}| H_0(T^{(s)}).$
It was shown~\cite{Man01} that for any $k \ge 0$ we have $H_k(T) \ge H_{k+1}(T)$
and $n H_k(T)$ is a lower bound to the output size of any compressor that encodes each symbol
of $T$ with a code that only depends on the symbol and its context of length $k$.

The technique of \emph{blocking}, i.e., to conceptually merge consecutive
symbols to form new symbols over a larger alphabet, 
is used to reduce the redundancy of Huffman encoding
for compressing a string.  The string $T$ of length $n$ is partitioned into $\frac{n}{\ell}$ blocks
of length $\ell$ each, then Huffman or other entropy codings are applied to compress a new string
$T_\ell$ of those blocks.  We call this operation \emph{blocking of length $\ell$}.

To prove our new results, we shall use the following theorem in
Section~\ref{sec:entropysimilar}:
\begin{theorem}[{\cite[Theorem 16.3.2]{CovTho91}}]\label{th:l1bound}
Let $p$ and $q$ be two probability mass functions on ${\cal A}$ such that
$|| p - q ||_1 \equiv \sum_{c \in {\cal A}} | p(c) - q(c) | \le \frac{1}{2}.$
Then
$|H_0(p) - H_0(q)| \le - || p - q ||_1 \log \frac{|| p - q ||_1}{|{\cal A}|}.$
\end{theorem}

\subsection{Review of Ferragina and Venturini's data structure}\label{sec:FV}

Here, we briefly review the data structure of Ferragina and
Venturini from~\cite{FerVen07b}.
It uses the same basic idea as Huffman coding:
replace every fixed-length block of symbols by a variable-length code
in such a way that frequently occurring blocks get shorter codes than
rarely occurring blocks.

To be more precise, consider a text $T[1..n]$ over an alphabet ${\cal A}$ 
where $|{\cal A}| = \sigma$ and $\sigma < n$.
Let $\ell = \frac{1}{2}\log_{\sigma} n$ and $\tau = \log n$.
Partition $T[1..n]$ into $\frac{n}{\tau \ell}$ super-blocks,
each contains $\tau \ell$ characters.
Each super-block is further partitioned into $\tau$ blocks,
each contains $\ell$ characters.
Denote the $\frac{n}{\ell}$ blocks
by $T_i = T[(i-1) \ell+1..i \ell]$ for $i=1, 2, \ldots, n/\ell$.

Since each block is of length $\ell$, there are 
at most $\sigma^{\ell} = \sqrt{n}$ distinct blocks.
For each block $P \in {{\cal A}}^{\ell}$,
let $f(P)$ be
the frequency of $P$ in $\{T_1, \ldots, T_{n/\ell} \}$.
Let $r(P)$ be the rank of $P$ according to the
decreasing frequency, i.e., the number of distinct blocks $P'$ such that $f(P') \ge f(P)$,
and $r^{-1}(j)$ be its inverse function.
Let $enc(j)$ be the rank $j$-th binary string in
$[\epsilon, 0, 1, 00, 01, 10, 11, 000, \ldots]$.

The data structure of Ferragina and Venturini consists of four arrays:
\begin{itemize}
\item[{\raise0.3pt\hbox{$\bullet$}}]
$V = enc(r(T_1)) \ldots enc(r(T_{n/\ell}))$.

\item[{\raise0.3pt\hbox{$\bullet$}}]
$r^{-1}(j)$ for $j=1, \ldots, \sqrt{n}$.

\item[{\raise0.3pt\hbox{$\bullet$}}]
Table $T_{Sblk}[1..\frac{n}{\ell \tau}]$ stores 
the starting position in $V$ of the encoding of every super-block.

\item[{\raise0.3pt\hbox{$\bullet$}}]
Table $T_{blk}[1..\frac{n}{\ell}]$ stores
the starting position in $V$ of the encoding of every block relative to 
the beginning of its enclosing super-block.
\end{itemize}

The algorithm for \textnormal{\texttt{access}}$(T,i)$ is simple:
Given $i$, compute the address where the block for $T[i]$ is encoded by
using $T_{Sblk}$ and $T_{blk}$ and obtain the code which encodes the rank
of the block.
Then, from $r^{-1}$, obtain the substring.
In total, this takes $\Order(1)$ time.
This yields:

\begin{lemma}[\cite{FerVen07b}]
Any substring $T[i..j]$ can be retrieved in
$\Order(1 + (j-i+1)/\log_{\sigma} n)$ time.
\end{lemma}

Using the data structure of Ferragina and Venturini,
$T[1..n]$ can be encoded using $n H_k + \Order(\frac{n}{\log_{\sigma} n} (k \log \sigma + \log \log n))$ bits according to the next lemma.

\begin{lemma}[\cite{FerVen07b}]\label{lem:FV}
The space needed by $V, r^{-1}, T_{Sblk}$, and $T_{blk}$ is as follows:
\begin{itemize}
\item[{\raise0.3pt\hbox{$\bullet$}}]
$V$ is of length $n H_k + 2 + \Order(k \log n) + \Order(n k \log \sigma / \ell)$ bits, simultaneously for all $0 \le k < \log_\sigma n$.
\item[{\raise0.3pt\hbox{$\bullet$}}]
$r^{-1}(j)$ for $j=1, \ldots, \sqrt{n}$ can be stored in
$\sqrt{n} \log n$ bits.
\item[{\raise0.3pt\hbox{$\bullet$}}]
$T_{Sblk}[1..\frac{n}{\ell \tau}]$ can be stored in $\Order(\frac{n}{\ell})$ bits.
\item[{\raise0.3pt\hbox{$\bullet$}}]
$T_{blk}[1..\frac{n}{\ell}]$ can be stored in $\Order(\frac{n}{\ell} \log\log n)$ bits.
\end{itemize}
\end{lemma}


\section{Entropies of similar strings}\label{sec:entropysimilar}

In this section, we prove that the empirical entropy of a string does not
change much after a small change to it.
This result will be used to bound the space complexity of our main
data structure in Section~\ref{sec:space}.
Consider two strings $T$ and $T^\prime$ of length $n$ and $n^\prime$,
respectively, such that the edit distance between $T$ and $T^\prime$ is one.
That is, $T^\prime$ can be obtained from $T$ by replacement, insertion, or deletion of one character.
We show that the empirical entropies of the two strings
do not differ so much.
\begin{theorem}\label{th:entropysimilar}
For two strings $T$ and $T^\prime$ on alphabet ${\cal A}$
of length $n$ and $n^\prime$ respectively,
such that the edit distance between $T$ and $T^\prime$ is one, and for any integer $k \ge 0$,
$\left| \ n H_k(T) - n^\prime H_k(T^\prime) \right| = \Order((k+1) (\log n + \log |{\cal A}|))$.
\end{theorem}

To prove Theorem~\ref{th:entropysimilar}, we first prove the following:

\begin{lemma}\label{lemma:1}
Let $T$ be a string of length $n$ over an alphabet ${\cal A}$, 
$T^-$ be a string made by deleting a character from $T$
at any position, $T^+$ be a string made by inserting a character into $T$
at any position,
and $T^\prime$ be a string by replacing a character of $T$ into another one
at any position.
Then the following relations hold:
\begin{eqnarray}
| nH_0(T) - (n-1)H_0(T^-) | &\le& 4 \log n + 3 \log |{\cal A}| \quad (\mbox{if $n \ge 1$}) \\
| nH_0(T) - (n+1)H_0(T^+) | &\le& 4 \log (n+1) + 4 \log |{\cal A}| \quad (\mbox{if $n \ge 0$}) \\
| nH_0(T) - nH_0(T^\prime) | &\le& 4 \log (n+1) + 3 \log |{\cal A}| \quad (\mbox{if $n \ge 0$})
\end{eqnarray}
\end{lemma}

\begin{proof}
Let $P(x)$, $P^-(x)$, $P^+(x)$, and $P^\prime(x)$ denote the empirical probability of a character $x \in {\cal A}$
in $T$, $T^-$, $T^+$, and $T^\prime$, respectively, and let $n_x$ denote the number of occurrences of $x \in {\cal A}$ in $T$.
It holds that $P(x) = \frac{n_x}{n}$ for any $x \in {\cal A}$.

If a character $c$ is removed from $T$, it holds that
$P^-(c) = \frac{n_c-1}{n-1}$, and $P^-(x) = \frac{n_x}{n-1}$
for any other $x \in {\cal A}$.  Then 
$||P - P^-||_1 = \frac{n-n_c}{n (n-1)}+ \sum_{x \in {\cal A}, x \neq c} \frac{n_x}{n (n-1)} = \frac{2(n-n_c)}{n (n-1)}$.
If $n = 1$, it holds $H_0(T) = 0$, and therefore $nH_0(T) - (n-1)H_0(T^-) = 0$ and the claim holds.
If $n = n_c$, which means that all characters in $T$ are $c$, it holds $H_0(T) = H_0(T^-) = 0$ and the claim holds.
Otherwise, $\frac{2}{n(n-1)} \le ||P - P^-||_1 \le \frac{2}{n}$ holds.
If $||P - P^-||_1 \le \frac{1}{2}$, from Theorem~\ref{th:l1bound}, 
$| H_0(P) - H_0(P^-) | \le - ||P - P^-||_1 \log \frac{||P - P^-||_1}{|{\cal A}|}
\le \frac{2}{n} \log \frac{|{\cal A}|n(n-1)}{2}$.
Then $|n H_0(T) - (n-1)H_0(T^-)| \le n | H_0(P) - H_0(P^-) | + H_0(P^-) \le 4 \log n + 3 \log |{\cal A}|$.
If $||P - P^-||_1 > \frac{1}{2}$, which implies $n < 4$, $|n H_0(T) - (n-1)H_0(T^-)| \le 3 \log |{\cal A}|$.
This proves the claim for $T^-$.

If a character $c$ is inserted into $T$, it holds that
$P^+(c) = \frac{n_c+1}{n+1}$, and $P^+(x) = \frac{n_x}{n+1}$
for any other $x \in {\cal A}$.  Then $||P - P^+||_1 = \frac{2(n-n_c)}{n (n+1)}$.
If $n = 0$, $H_0(T) = H_0(T^+) = 0$ and the claim holds.
If $n = n_c$, which means that $T^+$ consists of only the character $c$, $H_0(T) = H_0(T^+) = 0$ and the claim holds.
Otherwise, $\frac{2}{n(n-1)} \le ||P - P^+||_1 \le \frac{2}{n}$ holds.
If $||P - P^+||_1 \le \frac{1}{2}$, $|n H_0(T) - (n+1)H_0(T^+)| \le n | H_0(P) - H_0(P^+) | + H_0(P^-) \le 4 \log n + 3 \log |{\cal A}|$.
If $||P - P^+||_1 > \frac{1}{2}$, which implies $n < 4$, $|n H_0(T) - (n+1)H_0(T^+)| \le 4 \log |{\cal A}|$.
This proves the claim for $T^+$.

If a character $c$ of $T$ is replaced with another character $c^\prime \in {\cal A}$ ($c^\prime \neq c$),
$||P - P^\prime||_1 = \sum_{\alpha \in {\cal A}}
|P(\alpha) -  P^\prime(\alpha)| = |\frac{n_c}{n} - \frac{n_c-1}{n}|
+ |\frac{n_{c^\prime}}{n} - \frac{n_{c^\prime}+1}{n}| = \frac{2}{n}$.
If $||P - P^\prime||_1 \le \frac{1}{2}$,
$|n H_0(T) - nH_0(T^\prime)| \le n | H_0(P) - H_0(P^\prime) | \le 4 \log n + 2 \log |{\cal A}|$.
If $||P - P^\prime||_1 > \frac{1}{2}$, which implies $n < 4$, 
$|n H_0(T) - nH_0(T^\prime)| \le 3 \log |{\cal A}|$.
If $c^\prime = c$, $T^\prime = T$ and $|n H_0(T) - nH_0(T^\prime)| = 0$.
This completes the proof.
\qed
\end{proof}

By using this lemma, we prove the theorem.

\begin{proof}(of Theorem~\ref{th:entropysimilar})
From the definition of the empirical entropy, $n H_k(T) = \sum_{s \in {\cal A}^k} |T^{(s)}| H_0(T^{(s)})$.
Therefore for each context $s \in {\cal A}^k$, we estimate the change of $0$-th order entropy.

Because the edit distance between $T$ and $T^\prime$ is one,
these are expressed as $T = T_1 c T_2$ and $T^\prime = T_1 c^\prime T_2$
by two possibly empty strings $T_1$ and $T_2$, and possibly empty characters
$c$ and $c^\prime$.
For the context $T_1[n_1-k+1..n_1]$ ($n_1 = |T_1|$), denoted by $s_0$,
the character $c$ in the string $T^{(s_0)}$ will change to $c^\prime$.
The character $T_2[i]$ ($i=1,2,\ldots,k$) has the context $T_1[n_1-k+1+i..n_1] c T_2[1..i-1]$, denoted by $s_i$,
in $T$, but the context will change to $s^\prime_i = T_1[n_1-k+1+i..n_1] c^\prime T_2[1..i-1]$, in $T^\prime$.
Therefore a character $T_2[i]$ is removed from the string $T^{(s_i)}$, and it is inserted to $T^{\prime{(s_i)}}$.
Therefore in at most $2k+1$ strings ($T^{(s_0)}, T^{(s_1)}, \ldots, T^{(s_k)}, T^{\prime{(s_1)}}, \ldots, T^{\prime{(s_k)}}$),
the entropies will change.  From Lemma~\ref{lemma:1}, each one will change only $\Order(\log n + \log |{\cal A}|)$.
This proves the claim.
\qed
\end{proof}


\section{Memory management}\label{sec:memory}

This section presents a data structure for storing a set~$B$ of
$m$~variable-length strings over the alphabet~$\{0,1\}$,
which is an extension of the one
in~\cite{NavSad10}.
The data structure allows the contents of the strings and their lengths to
change, but the value of~$m$ must remain constant.
We assume a unit-cost word RAM model with word size~$w$ bits.
The memory consists of consecutively ordered bits, and any consecutive $w$ bits
can be accessed in constant time, as stated above.
A string over $\{0,1\}$ of length at most~$b$ is called
a \emph{$(\leq b)$-block}.
Our data structure stores a set~$B$ of $m$ such $(\leq b)$-blocks,
while supporting the following operations:
\begin{itemize}
\item[{\raise0.3pt\hbox{$\bullet$}}]
  \texttt{address}($i$):
  Return a pointer to where in the memory the $i$-th $(\leq b)$-block
  is stored ($1 \leq i \leq m$).
\item[{\raise0.3pt\hbox{$\bullet$}}]
  \texttt{realloc}($i,b'$):
  Change the length of the $i$-th $(\leq b)$-block to $b'$~bits
  ($0 \leq i \leq m$).
  The physical address for storing the block (\texttt{address}($i$)) may change.
\end{itemize}

\begin{theorem}
\label{theorem: Memory_management}
Given that $b \le m$ and $\log m \le w$,
consider the unit-cost word RAM model with word size $w$.
Let $B = \{B[1], B[2],$ $ \dots, B[m]\}$ be a set of $(\leq b)$-blocks
and let $s$ be the total number of bits of all $(\leq b)$-blocks in~$B$.
We can store $B$ in $s + \Order(m \log m + b^2)$ bits while supporting
\textnormal{\texttt{address}} in $\Order(1)$ time and
\textnormal{\texttt{realloc}} in $\Order(b/w)$ time.
\end{theorem}

\begin{theorem}
\label{theorem: Memory_management2}
Given a parameter $b = \Order(w)$,
consider the unit-cost word RAM model with word size $w$.
Let $B = \{B[1], B[2],$ $ \dots, B[m]\}$ be a set of $(\leq b)$-blocks,
and let
$s$ be the total number of bits of all $(\leq b)$-blocks in~$B$.
We can store $B$ in $s + \Order(w^4 + m \log w)$ bits while supporting
\textnormal{\texttt{address}} and \textnormal{\texttt{realloc}}
in worst-case $\Order(1)$ time.
\end{theorem}

\noindent
(Due to lack of space, the proofs of
Theorems~\ref{theorem: Memory_management}
and~\ref{theorem: Memory_management2}
are given in Appendix~\ref{sec:proof1}.)
From here on, we say that the data structure has parameters $(b, m)$.


\section{A data structure for maintaining the CRAM}
\label{sec:replace}

This section is devoted to proving Theorem~\ref{th:main1}.
Our aim is to dynamize Ferragina and Venturini's
data structure~\cite{FerVen07b} by allowing \texttt{replace} operations.
(Our data structure for the extended CRAM which also supports
\texttt{insert} and \texttt{delete} is described in Appendix~\ref{sec:indel}.)
Ferragina and Venturini's data structure uses a code table for encoding
the string, while our data structure uses two code tables, which will
change during update operations.

Given a string $T[1..n]$ defined over an alphabet ${\cal A}$ ($| {\cal A} | = \sigma$),
we support two operations. 
(1) $\texttt{access}(T,i)$: which returns $T[i..i+\frac{1}{2}\log_{\sigma} n-1]$; and 
(2) $\texttt{replace}(T, i, c)$: which replaces $T[i]$ with a character $c \in {\cal A}$.

We use blocking of length $\ell = \frac{1}{2}\log_{\sigma} n$ of $T$.
Let $T'[1..n']$ be a string of length $n' = \frac{n}{\ell}$ on an alphabet 
${\cal A}^\ell$
made by blocking of $T$.
The alphabet size is $\sigma^\ell = \sqrt{n}$.
Each character $T'[i]$ corresponds to the string $T[((i-1)\ell + 1)..i\ell]$.
A super-block consists of $1/\epsilon$ consecutive blocks in $T^\prime$
($\ell/\epsilon$ consecutive characters in $T$),
where $\epsilon$ is a predefined constant.

Our algorithm runs in phases.
Let $n'' = \epsilon n'$.
For every $j \ge 1$, we refer to the sequence of the $(n'' (j-1) + 1)$-th to
$(n'' j)$-th replacements as \emph{phase~$j$}.
The preprocessing stage corresponds to phase $0$.
Let $T^{(j)}$ denote the string just before phase $j$.  (Hence, $T^{(1)}$ is the input string $T$.)
Let $F^{(j)}$ denote the frequency table of blocks $b \in {\cal A}^\ell$ 
in $T^{(j)}$, 
and $C^{(j)}$ and $D^{(j)}$ a code table and a decode table defined below.
The algorithm also uses a bit-vector $R^{(j-1)}[1..n'']$,
where $R^{(j-1)}[i]=1$ means that the $i$-th super-block in $T$
is encoded by code table $C^{(j-1)}$; otherwise, it is encoded
by code table $C^{(j-2)}$.

During the execution of the algorithm, we maintain the following invariant:
\begin{itemize}
\item[{\raise0.3pt\hbox{$\bullet$}}]
At the beginning of phase $j$, the string $T^{(j)}$ is encoded with
code table $C^{(j-2)}$ (we assume $C^{(-1)} = C^{(0)} = C^{(1)}$), and
the table $F^{(j)}$ stores the frequencies of blocks in $T^{(j)}$.
\item[{\raise0.3pt\hbox{$\bullet$}}]
During phase $j$, the $i$-th super-block is encoded with
code table $C^{(j-2)}$ if $R^{(j-1)}[i]=0$, or $C^{(j-1)}$ if $R^{(j-1)}[i]=1$.
The code tables $C^{(j-2)}$ and $C^{(j-1)}$ do not change.%
\item[{\raise0.3pt\hbox{$\bullet$}}]
During phase $j$, $F^{(j+1)}$ stores the correct frequency of blocks
of the current~$T$.
\end{itemize}

\subsection{Phase $0$: Preprocessing}

First, for each block $b \in {\cal A}^\ell$,
we count the numbers of its occurrences in $T^\prime$
and store it in an array $F^{(1)}[b]$.
Then we sort the blocks $b \in {\cal A}^\ell$ in decreasing order of
the frequencies $F^{(1)}[b]$, and assign a code $C^{(1)}[b]$ to encode them.
The code for a block $b$ is defined as follows.
If the length of the code $enc(b)$, defined in Section~\ref{sec:FV},
is at most $\frac{1}{2}\log n$ bits, then $C^{(1)}[b]$ consists of a bit `0',
followed by $enc(b)$.  Otherwise, it consists of a bit `1',
followed by the binary encoding of $b$, that is, the block is stored without
compression.
The code length for any block $b$ is upper bounded 
by $1+\frac{1}{2}\log n$ bits.
Then we construct a table $D^{(1)}$ for decoding a block.
The table has $2^{1+\frac{1}{2}\log n} = \Order(\sqrt{n})$ entries
and $D^{(1)}[x] = b$ for all binary patterns $x$ of length $1+\frac{1}{2}\log n$
such that a prefix of $x$ is equal to $C^{(1)}[b]$.  Note that this decode table
is similar to $r^{-1}$ defined in Section~\ref{sec:FV}.

Next, for each block $T^\prime[i]$ ($i=1,\ldots,n'$), 
compute its length using $C^{(1)}[T^\prime[i]]$,
allocate space for storing it using the data structure of Theorem~\ref{theorem: Memory_management2}
with parameters 
$(1+\ell \log\sigma ,\frac{n}{\ell}) = (1+\frac{1}{2}\log n, \frac{2n \log\sigma}{\log n})$,
and $w = \log n$.
From Lemma~\ref{lem:FV} and Theorem~\ref{theorem: Memory_management2},
if follows that
the size of the initial data structure is $nH_k(T) + \Order\left(\frac{n\log\sigma}{\log n}
(k \log\sigma + \log\log n)\right)$
bits.
Finally, for later use, copy the contents of $F^{(1)}$ to $F^{(2)}$,
and initialize $R^{(0)}$ by $0$.
By sorting the blocks by a radix sort, the preprocessing time becomes
$\Order(n\log\sigma/\log n)$.

\subsection{Algorithm for \texttt{access}}
The algorithm for \texttt{access}$(T,i)$ is:
Given the index $i$,
compute the block number $x = \lfloor (i-1)/\ell \rfloor +1$
and the super-block number $y$ containing $T[i]$.
Obtain the pointer to the block and the length of the code
by \texttt{address}($x$).
Decode the block using the decode table $D^{(j-2)}$ if $R^{(j-1)}[x]=0$,
or $D^{(j-1)}$ if $R^{(j-1)}[x]=1$.
This takes constant time.

\subsection{Algorithm for \texttt{replace}}
We first explain a naive, inefficient algorithm.
If $b = T^\prime[i]$ is replaced with $b^\prime$, we change the frequency
table $F^{(1)}$ 
so that $F^{(1)}[b]$ is decremented by one and $F^{(1)}[b^\prime]$ is
incremented by one.  Then new code table $C^{(1)}$ and decode table $D^{(1)}$
are computed from updated $F^{(1)}$,
and all blocks $T^\prime[j]$ ($j=1,\ldots,n'$) are re-encoded by 
using the new code table.
Obviously, this algorithm is too slow.

To get a faster algorithm, we can delay updating code tables for the
blocks and re-writing the blocks using new code tables
because of Theorem~\ref{th:entropysimilar}.  Because the amount of change
in entropy is small after a small change in the string,  we can show that
the redundancy of using code tables defined according to an old string can be
negligible.  
For each single character change in $T$, we re-encode
a super-block ($\ell/\epsilon$ characters in $T$).
After $\epsilon n'$ changes, the whole string will be re-encoded.
To specify which super-block to be re-encoded,
we use an integer array $G^{(j-1)}[1..n'']$.
It stores a permutation of
$(1,\ldots,n'')$ and indicates
that at the $x$-th replace operation in phase $j$
we rewrite the $G^{(j-1)}[x]$-th super-block.
The bit $R^{(j-1)}[x]$ indicates 
if the super-block has been already rewritten or not.
The array $G^{(j-1)}$ is defined by sorting super-blocks
in increasing order of lengths of codes for encoding super-blocks.

We implement $\texttt{replace}(T, i, S)$ as follows.
In the $x$-th update in phase~$j$,
\begin{enumerate}
\item
If $R^{(j-1)}[G^{(j-1)}[x]]=0$, i.e., 
if the $G^{(j-1)}[x]$-th super-block is encoded with $C^{(j-2)}$,
decode it and re-encode it with $C^{(j-1)}$, and
set $R^{(j-1)}[G^{(j-1)}[x]] = 1$.
\item
Let $y$ be the super-block number containing $T[i]$,
 that is, $y = \lfloor \epsilon (i-1) / \ell \rfloor$.
\item
Decode the $y$-th super-block, which is encoded with $C^{(j-2)}$ or
$C^{(j-1)}$ depending on $R^{(j-1)}[y]$.
Let $S'$ denote the block containing $T[i]$.
Make a new block $S$ from $S'$ by applying the \texttt{replace} operation.
\item
Decrement the frequency $F^{(j+1)}[S']$ 
and increment the frequency $F^{(j+1)}[S]$.
\item
Compute the code for encoding $S$ using $C^{(j-1)}$ if the $y$-th super-block
is already re-encoded ($R^{(j-1)}[y]=1$), or $C^{(j-2)}$ otherwise ($R^{(j-1)}[y]=0$).
\item
Compute the lengths of the blocks in $y$-th super-block and apply
\texttt{realloc} for those blocks.
\item
Rewrite the blocks in the $y$-th super-block.
\item
Construct a part of tables $C^{(j)}$, $D^{(j)}$, $G^{(j)}$, and $R^{(j)}$
(see below).
\end{enumerate}

To prove that the algorithm above maintains the invariant, we need only to prove
that the tables $C^{(j-1)}$, $F^{(j)}$, and $G^{(j-1)}$
are ready at the beginning of phase $j$.
In phase $j$, we create $C^{(j)}$ based on $F^{(j)}$.  
This is done by just radix-sorting
the frequencies of blocks, and therefore the total time complexity is
$\Order(\sigma^l) = \Order(\sqrt{n})$.
Because phase $j$ consists of 
$n''$ \texttt{replace} operations, the work for creating $C^{(j)}$
can be distributed in the phase.
We represent the array $G^{(j-1)}$ implicitly by 
$(1/\epsilon)(1+\frac{1}{2}\log n)$ doubly-linked lists $L_d$; 
$L_d$ stores super-blocks of length $d$.  By retrieving the lists 
in decreasing order of $d$ we can
enumerate the elements of $G^{(j)}$.  If all the elements of a list have been
retrieved, we move to the next non-empty list.  This can be done in 
$\Order(1/\epsilon)$ time if we use a bit-vector of
$(1/\epsilon)(1+\frac{1}{2}\log n)$ bits indicating which lists are non-empty.
We copy $F^{(j)}$ to $F^{(j+1)}$ in constant time
by changing pointers to $F^{(j)}$ and $F^{(j+1)}$.
For each \texttt{replace} in phase $j$, 
we re-encode a super-block, which consists of
$1/\epsilon$ blocks.  This takes $\Order(1/\epsilon)$ time.
Therefore the time complexity for \texttt{replace} is $\Order(1/\epsilon)$ time.

Note that during phase $j$, only the tables
$F^{(j)}$, $F^{(j+1)}$, 
$C^{(j-2)}$, $C^{(j-1)}$, $C^{(j)}$, 
$D^{(j-2)}$, $D^{(j-1)}$, $D^{(j)}$, 
$G^{(j-1)}$, $G^{(j)}$, 
$R^{(j-1)}$, and $R^{(j)}$
are stored.
The other tables are discarded.

\subsection{Space analysis}\label{sec:space}

Let $s(T)$ denote the size of the encoding of $T$ by our dynamic data structure.
At the beginning of phase $j$, the string $T^{(j)}$ is encoded with
code table $C^{(j-2)}$, which is based on the string $T^{(j-2)}$.
Let $L^{(j)} = nH_k(T^{(j)})$ and $L^{(j-2)} = nH_k(T^{(j-2)})$.

After the preprocessing,
$s(T^{(1)}) \le L^{(1)} + \Order\left(\frac{n\log\sigma}{\log n}(k\log\sigma+\log\log n)\right)$.
If we do not re-encode the string, for each \texttt{replace} operation we write at most
$1+\frac{1}{2}\log n$ bits.  Therefore $s(T^{(j)}) \le s(T^{(j-2)}) +\Order(n''\log n)$ holds.
Because $T^{(j)}$ is made by $2(n''+\sqrt{n})$ character changes to $T^{(j-2)}$,
from Theorem~\ref{th:entropysimilar}, we have
$ |L^{(j)} - L^{(j-2)}| = \Order(n'' (k+1)(\log n+\log\sigma))$.
Therefore we obtain
$s(T^{(j)}) \le L^{(j)} + \Order(\epsilon(k+1)n\log\sigma)$.
The space for storing the tables $F^{(j)}$, $C^{(j)}$, $D^{(j)}$, $G^{(j)}$,
$H^{(j)}$, and $R^{(j)}$ is
$\Order(\sqrt{n}\log n)$, $\Order(\sqrt{n}\log n)$, $\Order(\sqrt{n}\log n)$,
$\Order(n'' \log n) = \Order(\epsilon n \log\sigma)$,
$\Order(n'' \log n)$,
$\Order(n'')$ bits, respectively.

Next we analyze the space redundancy caused by the re-encoding of super-blocks.
We re-encode the super-blocks with the new code table in increasing order of their lengths,
that is, the shortest one is re-encoded first,
This guarantees that
at any time the space does not exceed $\max\{s(T^{(j)}), s(T^{(j-2)})\}$.
This completes the proof of Theorem~\ref{th:main1}.


\section{Concluding remarks}\label{sec:conclusion}

We have presented a data structure called Compressed Random Access Memory
(CRAM), which compresses a string $T$ of length $n$ into its $k$-th order
empirical entropy in such a way that any consecutive $\log_\sigma n$ bits
can be obtained in constant time (the \texttt{access} operation), and
replacing a character (the \texttt{replace} operation) takes
$\Order(\min\{\log_\sigma n, (k+1)\log n/\log\log n\})$ time.
The time for \texttt{replace} can be reduced to constant
($\Order(1/\epsilon)$) time
by allowing an additional $\Order(\epsilon(k+1) n\log\sigma)$ bits redundancy.
The extended CRAM data structure also supports the \texttt{insert} and
\texttt{delete} operations, at the cost of increasing the time for
\texttt{access} to $\Order(\log n/\log \log n)$ time, which is optimal under
this stronger requirement, and the time for each update operation also
becomes $\Order(\log n/\log \log n)$.

An open problem is how to improve the running time of \texttt{replace}
for the CRAM data structure to $\Order(1)$ without using the
$\Order(\epsilon(k+1) n\log\sigma)$ extra bits.


\subsection*{Acknowledgments}
JJ~was funded by the Special Coordination Funds for Promoting Science and
Technology, Japan.
KS~was supported in part by KAKENHI 23240002.
WKS~was supported in part by the MOE's AcRF Tier 2 funding
R-252-000-444-112.


\bibliographystyle{plain}
\bibliography{Bibl_compressed_RAM}


\newpage

\appendix

\section*{Appendix}

\section{Memory management}\label{sec2:memory}\label{sec:proof1}

This appendix proves Theorems~\ref{theorem: Memory_management}
and~\ref{theorem: Memory_management2}
from Section~\ref{sec:memory}.

\medskip

We first describe some technical details.
(Recall the definitions from Section~\ref{sec:memory}.)
Let $p = \Theta(\log (mb))$.
This is the number of bits needed to represent a memory address
relative to the head of a memory region storing $B$.
For any $(\leq b)$-block~$B[i]$ in~$B$, we will refer to the current contents
of~$B[i]$ by~$data(i)$.
To store $data(i)$ for all $(\leq b)$-blocks compactly while allowing efficient
updates, we allocate memory in such a way that $data(i)$ for any given
$(\leq b)$-block~$B[i]$ may be spread out over at most two non-consecutive regions in the
memory.
For this purpose, define a \emph{segment} to be $b + 4p$~consecutive bits
of memory.
The data structure will always allocate and deallocate memory in terms of
segments, and keep a pointer to the start of the segment currently at the
highest address in the memory.

The core of our data structure is $b$ doubly-linked lists
$L_{1},\, \dots,\, L_{b}$, where
each list $L_x$ is a doubly-linked list of a set of segments,
each is of length $b+4p$ bits.
The main idea is to use the segments in list~$L_{x}$ to store all
$(\leq b)$-blocks of length~$x$.
Moreover, we do so in such a way that when the length of a
$(\leq b)$-block changes from~$x$ to~$x+d$, we only need to update a
few segments in the two lists~$L_{x}$ and~$L_{x+d}$.

Every segment belonging to a list~$L_{x}$ is used as follows:
\begin{itemize}
\item[{\raise0.3pt\hbox{$\bullet$}}]
  \emph{pred}:
  $p$~bits to store the memory address of its predecessor segment in~$L_{x}$.
\item[{\raise0.3pt\hbox{$\bullet$}}]
  \emph{succ}:
  $p$~bits to store the memory address of its successor segment in~$L_{x}$.
\item[{\raise0.3pt\hbox{$\bullet$}}]
  \emph{block\_data}:
  $b+p$~bits for storing information associated with one or more $(\leq b)$-blocks
  of length~$x$.
  More precisely, for $i \in [m]$, let $id(i)$ denote the binary
  encoding of the integer~$i$ in $\log m \le p$~bits.
  For every $x \in [b]$, form a string~$B^{\ast}_{x}$ by
  concatenating the pairs~$(id(i), data(i))$ for all $(\leq b)$-blocks~$B[i]$ of
  length~$x$ in some arbitrary order.
  Divide $B^{\ast}_x$ into substrings of length~$b$ and store them in the
  \emph{block\_data} region of consecutive segments in~$L_{x}$.
  Only the first segment in each list~$L_{x}$ is allowed to have some unused
  bits in its \emph{block\_data} region.
\item[{\raise0.3pt\hbox{$\bullet$}}]
  \emph{offset}:
  $\log b \, (\leq p)$~bits to store the relative starting position within
  the \emph{block\_data}-region of the first $(\leq b)$-block.
  Observe that the head \emph{offset} bits of the segment are used
  to store another $(\leq b)$-block, whose starting position belongs to another
  segment, except the first segment in $L_{x}$.
\end{itemize}

\begin{figure}[h!]
\begin{center}
  \includegraphics[scale=0.35]{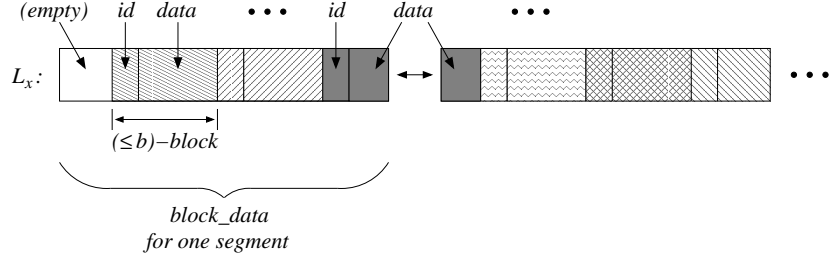}
\caption{Each doubly-linked list~$L_{x}$ consists of segments which together
store all $(\leq b)$-blocks of length~$x$ in their \emph{block\_data}-regions
(shown here).
The data for a single $(\leq b)$-block may be split across two segments
and the first segment in~$L_{x}$ may contain some unused bits.}
\label{figure2: fig_memory_management_B}
\vspace*{-5mm}
\end{center}
\end{figure}

See Fig.~\ref{figure2: fig_memory_management_B} for an illustration.
Note that a $(\leq b)$-block~$B[i]$ may stretch across two segments
in~$L_{x}$, and that these two segments might not be located in
a consecutive region of the memory.

In addition to the doubly-linked lists, we use four arrays
$Seg[1..m]$, $Pos[1..m]$, $Len[1..m]$, and $Ind[1..m]$ to
remember the locations of the $(\leq b)$-blocks.
To be precise, for each $i \in [m]$, we have
\begin{itemize}
\item[{\raise0.3pt\hbox{$\bullet$}}]
$Seg[i]$ contains a pointer $j$ to $Ind[j]$.
\item[{\raise0.3pt\hbox{$\bullet$}}]
$Pos[i]$ contains the relative location of $B[i]$
from the starting position of the segment.
\item[{\raise0.3pt\hbox{$\bullet$}}]
$Len[i]$ stores the length of $B[i]$.
\item[{\raise0.3pt\hbox{$\bullet$}}]
$Ind[j]$ stores the memory address of the segment which stores blocks
$B[i]$ with $Seg[i] = j$.
\end{itemize}
The array $Ind$ is used for indirect addressing.
Consider the case that we do not use the array $Ind$ and store memory addresses 
in $Seg$ directly.
When a segment has moved to another address, we have to rewrite non-constant
number of $Seg[i]$'s for $i$'s such that the $i$-th block is stored in the segment.
On the other hand, if $Ind$ is used,  we need not to rewrite all $Seg[i]$'s and
it is enough to rewrite only an entry $Ind[j]$ ($j = Seg[i]$ for those $i$'s).

\medskip

\begin{proof} (of Theorem~\ref{theorem: Memory_management})
To implement \texttt{address}$(i)$ in $\Order(1)$ time, 
we simply return three values $Q = Seg[Ind[i]]$, $Pos[i]$, and $Len[i]$.
Observe that $B[i]$ is stored at position $Pos[i]$ of the segment $Q$.
Moreover, when $Pos[i]+Len[i]>b+p$, the $(\leq b)$-block $B[i]$ spans
across two segments (i.e. segments $Q$ and $Q.\emph{succ}$).
In this case, the first $Pos[i]+Len[i]-b-p$ bits of $B[i]$ are
stored at the end of segment $Q$ while
the remaining bits of $B[i]$ are stored
at the beginning of segment $Q.\emph{succ}$.

To implement \texttt{realloc}$(i,b')$ in $\Order(b/w)$ time, we first find the
location~$q$ of $B[i]$ by \texttt{address}($i$).
Store $data(i)$ in a temporary space~$tmp$.
Obtain the index $j$ of the first $(\leq b)$-block~$B[j]$ in~$L_{x}$
from the \emph{block\_data} of the first segment in~$L_{x}$.
Copy the first $(\leq b)$-block~$B[j]$ in~$L_{x}$ to position~$q$ and
update~$Seg[j]$ and $Pos[j]$ accordingly.
If the first segment in~$L_{x}$ becomes empty, move the segment
with the largest address to the empty location.
Let $s$ and $t$ be the addresses of the segment before and after the movement,
respectively.
We have to change pointers to all $(\leq b)$-blocks $B[j']$
stored in the moved segment.  Though the moved segment may contain $\Theta(b)$
$(\leq b)$-blocks, this can be done in constant time as follows.
For any $(\leq b)$-block $B[j']$ in the segment, it holds $Ind[j'] = r$ for some
$r$, and $Seg[r] = s$ before the movement.
Therefore it holds $Seg[Ind[j']] = s$ for all the blocks.
After the movement, it must hold that $Seg[Ind[j']] = t$ for all the blocks.
This is done in constant time by simply setting $Seg[r] = t$ 
and we need not change $Ind[j']$ for each $j'$.
Each value $r = Ind[j']$ is an integer in the range $[0,m]$ and corresponds to
a segment.  Precisely,  if the data structure currently has $r$ segments, and
a new segment is allocated for $B[j]$, we set $Ind[j] = r+1$ and $Seg[r+1]$
is set to the address of the new segment.  If the blocks in the $(r+1)$-st segment
have been moved, we change $Seg[r+1]$.  

Next, copy the $x$~bits stored in~$tmp$ to immediately before the first
$(\leq b)$-block in the head of the list~$L_{x+d}$.
In case the first segment of~$L_{x+d}$ does not have enough space to store
it, we allocate a new segment.
Finally, update $Seg[i]$, $Pos[i]$, $Len[i]$, and $Ind[i]$.
The time complexity is $\Order(\frac{b}{w})$.  This is necessary to copy
segments of $\Order(b+p)$ bits.

\medskip

To analyze the space complexity, note that the total length of $data(i)$ for
all $B[i] \in B$ is~$s$.
Each list $L_x$ has at most one segment which has an empty slot, and therefore
there exist at most $b$ segments with an empty slot.
Then the number of segments to store $s$ bits is at most $s/b+b$ and
the space to store the segments is 
$(b+4p)(s/b+b) = s(1+\frac{4p}{b}) + b(b+4p)$.
This is $s + \Order(b^2 + m \log m)$ because $p = \log(mb) = \Order(\log m)$
and $s \le bm$.
We also need $4mp = \Order(m \log m)$ bit space 
for storing $Seg[1..m]$, $Ind[1..m]$, $Pos[1..m]$, and $Len[1..m]$.
In total, the total space is
$s + \Order(b^2 + m \log m)$ bits.
This proves Theorem~\ref{theorem: Memory_management}.
\qed
\end{proof}

The data structure of Theorem~\ref{theorem: Memory_management} uses
$\Order(m \log m)$ bits additional space, which is too large if we want
to store many short blocks.
Because of this, we give an alternative data structure in
Theorem~\ref{theorem: Memory_management2}.
We say the data structure has parameters $(b, m)$.

\begin{proof}  (of Theorem~\ref{theorem: Memory_management2})
We use Theorem~\ref{theorem: Memory_management} to store a set of $(\leq b)$-blocks
$B[1..m]$ for $b = 1+w$ and $m = w^c$ for some constant $c>0$.
We say the data structure has parameters $(1+w, w^c)$.
For general $m > w^c$, we split $B[1..m]$ into $m' = \lceil m/w^c \rceil$ sub-arrays
$B'_1[1..w^c], B'_2[1..w^c], \ldots, B'_{m'}[1..w^c]$ and to store each sub-array $B'_i$
we use Theorem~\ref{theorem: Memory_management} with parameters $(b,m) = (1+w, w^c)$.
Then \texttt{address}($i$) and \texttt{realloc}($i,b$) are done in constant time.
However a problem is that the memory space to store the data structure 
for each sub-array will change, and we cannot store it in a consecutive memory region.
To overcome this, we use a two-level data structure.
Let $M = \lceil \frac{m}{w^3} \rceil$.
The higher level consists of $M$ data structures of 
Theorem~\ref{theorem: Memory_management} with parameters $(1+w, w^3)$.
We call each one $D_i$ 
($i = 1,\ldots,M$).  Each $D_i$ uses a consecutive
memory region, which is impossible.  Therefore we use a kind of \emph{virtual memory}.
The memory to store segments is divided into pages.
Each page uses a physically consecutive memory region, while the pages are
located in non-consecutive regions.
Each data structure $D_i$ has $w^2$ pages, and each page contains either $w$ segments
or no segments, depending on how many bits are necessary to store the blocks.
Precisely, if $s$ segments are necessary, the first $\lceil s/w \rceil$ pages
have $w$ segments each, and the rest have no segments.
The total number of pages for all $D_i$ ($i = 1,\ldots,M$) is 
$M w^2 = \lceil \frac{m}{w} \rceil$.

The whole pages of all the data structures $D_i$ ($i = 1,\ldots,M$)
are managed by a single data structure of Theorem~\ref{theorem: Memory_management}
with parameters $(\Order(w^2), \lceil \frac{m}{w} \rceil)$.
We call the data structure $D$.  
The algorithm for \texttt{address}($i$) becomes as follows.
Let $q = \lfloor \frac{i-1}{w^3} \rfloor + 1$ and $r = i - (q-1)w^3$.
The $i$-th block is stored as the $r$-th block of $D_q$.
Therefore we compute \texttt{address}($r$) in $D_q$, and obtain the logical (virtual)
address $x$ of the segment containing the block.
To convert it into the physical (real) address $y$ of the memory,
we first compute the page number $z$ for the segment, then compute
\texttt{address}($z$) in $D$.  It is straightforward to compute the address
for the block inside the page because the segments in the page are of
the same length.
These operations are done in constant time.

The function \texttt{realloc}($i,b$) is implemented as follows.
First we find the data structure $D_q$ that contains the $i$-th block,
and execute \texttt{realloc} in $D_q$.  Note that $D_q$ uses virtual memory
which is managed by $D$.  Therefore we have to convert a logical address to
a physical one for any memory access.
During the execution of \texttt{realloc} in $D_q$, we have to move a constant
number of pages of $w^2$ bits in $D$.
If we naively use the result of Theorem~\ref{theorem: Memory_management},
it takes $\Order(w)$ time.  It is easy to obtain an amortized $\Order(1)$ time
algorithm because each block is of $\Order(w)$ bits and movement of pages in $D$
occurs every $\Order(w)$ operations of \texttt{realloc} in $D_q$.
It is also easy to obtain a worst-case $\Order(1)$ time algorithm at the cost
of $\Order(w^2)$ bit redundant space for each $D_q$.  We can spread the movement
of pages over the next $\Order(w)$ execution of \texttt{realloc} in $D_q$.
Then a constant number of pages with empty slots will exist, resulting in $\Order(w^2)$
bit redundancy in space.  This redundancy sums up to 
$\Order(\frac{m}{w^3}\cdot w^2) = \Order(\frac{m}{w})$ bits for all $D_q$.

We analyze the space complexity.  Each $D_q$ has $\Order(w^2 + w^3 \log w)$-bit
auxiliary data structures, and they sum up to $\Order(m \log w)$ bits.
The data structure $D$ uses $s + \Order(w^4 + \frac{m \log w}{w})$ bits.
Therefore the total space is $s + \Order(w^4 + m \log w)$ bits.
\qed
\end{proof}


\section{A data structure for maintaining the extended CRAM}
\label{sec:indel}

This section proves Theorem~\ref{th:main2}.
To support \texttt{insert} and \texttt{delete} efficiently,
we use variable length super-blocks.  
Namely, let $\tau = \frac{\log n}{\log \log n}$ and $\ell = \frac{1}{2}\log_{\sigma} n$.
Each super-block consists of $\tau$ to $2\tau$ blocks
($\tau\ell$ to $2\tau\ell$ consecutive characters in $T$).
These super-blocks are stored using the data structure of Theorem~\ref{theorem: Memory_management} with parameters $(2\log^2 n/\log\log n, n\log\log n/\log^2 n)$.
To represent super-block boundaries, we use a bit-vector $B[1..n]$
such that $B[i]=1$ means that $T[i]$ is the first character in a super-block.
Therefore $B$ has $\Theta(n \log\sigma \log\log n/\log^2 n)$ ones.  
This bit-vector is stored using the following data structure.
\begin{lemma}[Lemma 17~\cite{NavSad10}]\label{lem2:GS}
We can maintain any bit-vector $B[1..n]$ within $nH_0(B)+\Order(n \log \log n/$ $\log n)$ bits of space,
while supporting the operations $\textnormal{\texttt{rank}}$, $\textnormal{\texttt{select}}$,
 $\textnormal{\texttt{insert}}$, and $\textnormal{\texttt{delete}}$,
all in time $\Order(\log n/\log \log n)$.
\end{lemma}

Because $nH_0(B) = \Order(n\log\log n/\log n)$, the bit-vector can be stored
in $\Order(n\log\log n/\log n)$ bits,
and $\texttt{rank}(B,i)$ can be computed in $\Order(\log n/\log \log n)$ time,
where $\texttt{rank}(B,i)$ is the number of ones in $B[1..i]$.
By using this data structure, we can compute the super-block number containing $T[i]$ by $\texttt{rank}(B,i)$.

The algorithm for \texttt{insert} and \texttt{delete} in
the extended CRAM works as follows.
First, we re-write the super-block in which \texttt{insert} or \texttt{delete} occurs.
If it contains less than $\tau$ or more than $2\tau$ blocks, we merge two consecutive super-blocks
or split it into two to maintain the invariant that every super-block consists of $\tau$ to $2\tau$ blocks.  If the lengths of super-blocks change, we update the bit-vector $B$ accordingly.
Because only a constant number of super-blocks change, the time complexity is 
$\Order(\log n/\log \log n)$.

However, in the extended CRAM, the time complexity of $\texttt{access}(T,i)$
increases from $\Order(1)$ to $\Order(\log n/\log \log n)$:
We first compute the super-block number for $T[i]$ by using
$\texttt{rank}(B,i)$ in $\Order(\log n/\log \log n)$ time.
Then, we scan all the blocks in the super-block
to find the location storing $T[i..i+\ell-1]$.
Furthermore, for each $\texttt{replace}(T,i,c)$ operation, we re-encode
a super-block.
This means that we set $\epsilon = \Order(\log\log n/\log n)$ in Theorem~\ref{th:main1}.
The time complexity becomes $\Order(\log n/\log \log n)$.

We also have to consider ``the change of $\log n$ problem''.
The sizes of blocks and super-blocks
depend on $\log n$, the length of the string.  If $n$ changes a lot, $\log n$ also changes, and
we have to reconstruct the data structure for the new value of $\log n$.  To avoid this,
we use the same technique as M{\"a}kinen and Navarro~\cite{MNtalg08}.  We partition the string $T$ into three parts
$T_1, T_2, T_3$, and encode them using $\log n -1, \log n, \log n+1$ as their ``$\log n$'' values,
respectively.  
We maintain the following invariant that
if $n$ is zero or a power of two, $T_1$ and $T_3$ are empty and $T_2$ is
equal to $T$, and
if $n$ increases by one, the length of $T_3$ grows by two and that of $T_1$
shrinks by one.  To accomplish this,
depending on where an insert occurs, we move the rightmost one or two characters
of $T_1$ to the beginning of $T_2$, and the rightmost one or two characters of $T_2$ to the beginning of $T_3$.
A deletion is done similarly.
We can guarantee that if the string length is doubled,
``$\log n$'' increases by one.
For example, if $n$ is a power of two then all the characters belong to $T_2$.
Then after $n$ insertions, the length becomes $2n$ and
all the characters move to $T_3$, and now $T_3$ becomes the new $T_2$.
The strings $T_1$, $T_2$, and $T_3$ are stored using
the data structure of Theorem~\ref{theorem: Memory_management} with parameters
$(2(\log n-1)^2/\log(\log n-1), n\log(\log n-1)/(\log n-1)^2)$,
$(2(\log n)^2/\log\log n, n\log\log n/(\log n)^2)$,
$(2(\log n+1)^2/\log(\log n+1), n\log(\log n+1)/(\log n+1)^2)$,
respectively.
The asymptotic space and time complexities do not change.


\section{Experimental results}
\label{sec:experiments}

To test the performance of the CRAM data structure in practice, we
implemented a prototype named \texttt{cram} and compared it to
a \texttt{gzip}-based alternative method named \texttt{cramgz}.
For the implementations, we used the C~programming language.
In this appendix, we describe the details of the experiments and the outcome.

\begin{enumerate}
\item
\texttt{cram} is the data structure that we introduced in this paper, but
a few changes were made to make it easier to implement.
(These simplifications may reduce the performance of the method slightly;
however, as shown below, it is still excellent.)
To be precise, the \texttt{cram}-prototype works as described in the paper
but with the following modifications:

\smallskip

\begin{itemize}
\item[{\raise0.3pt\hbox{$\bullet$}}]
The memory is partitioned into \emph{large blocks}, and the size of each
large block is denoted by~$b$.
Each large block is further partitioned into \emph{middle blocks} of
length~$m$.
Finally, each middle block stores \emph{blocks} of length
$\ell = 2$, instead of $\frac{1}{2}\log_\sigma n$, 
which correspond to the ``blocks'' described in the paper.

\smallskip
\item[{\raise0.3pt\hbox{$\bullet$}}]
Pointers to large blocks and middle blocks are stored.

\smallskip
\item[{\raise0.3pt\hbox{$\bullet$}}]
Each block is encoded by a Huffman code.
We assign codes to all characters in the alphabet because otherwise we
cannot encode characters that are missing from the initial string but later
appear due to \texttt{replace} operations.

\smallskip
\item[{\raise0.3pt\hbox{$\bullet$}}]
An additional parameter~$u$ is employed as follows.
If $x$ bytes of memory change occur, then $x \cdot u$ bytes are re-encoded
with new codes.
Furthermore, if $x \cdot u > b$, a large block is re-encoded.
(This means that the worst-case time complexity of \texttt{replace}
becomes amortized, so it is no longer constant.)
\end{itemize}

\medskip

\item
Next, \texttt{cramgz} is an original data structure based on \texttt{gzip}:

\smallskip

\begin{itemize}
\item[{\raise0.3pt\hbox{$\bullet$}}]
The memory is partitioned into \emph{large blocks} of length~$b'$.

\smallskip
\item[{\raise0.3pt\hbox{$\bullet$}}]
Each large block is compressed by \texttt{gzip} and stored using our
dynamic memory management algorithm.
We employed the zlib library\footnote{\texttt{http://zlib.net}}, with
compression level~$1$ (i.e., the fastest compression).
\end{itemize}
\end{enumerate}

\noindent
For the experiments, we used the two text files \emph{English} and \emph{DNA}
from the Pizza \& Chili
corpus~\footnote{\texttt{http://pizzachili.dcc.uchile.cl}}.
The lengths of these texts are limited to 200MB, so $n = 200 \cdot 2^{20}$
and the alphabet size $\sigma = 256$.

\bigskip

\noindent
\textbf{Experiment~1}:
The first experiment only involved \texttt{cram} and was done to determine
a suitable value for the parameter~$u$.
We first built the CRAM data structure for the English text using the
parameters $b = 1024$, $m = 64$.
Then, we overwrote the DNA text onto the (compressed) English text one
character at a time, from left to right, by using \texttt{replace}
operations.
The $1$-st order entropy of the initial text was about $4$~bits per
character (bpc), and including the auxiliary data structure, the size was
about $4.67$~bpc.
On the other hand, since the DNA text only consists of four distinct
characters $\{$a, c, g, t$\}$ and the text is almost random,
its $1$-st order entropy is $2$~bpc and
about $2.67$~bpc including the auxiliary data structure.

\begin{figure}[t!]
\begin{center}
  \includegraphics[scale=0.90]{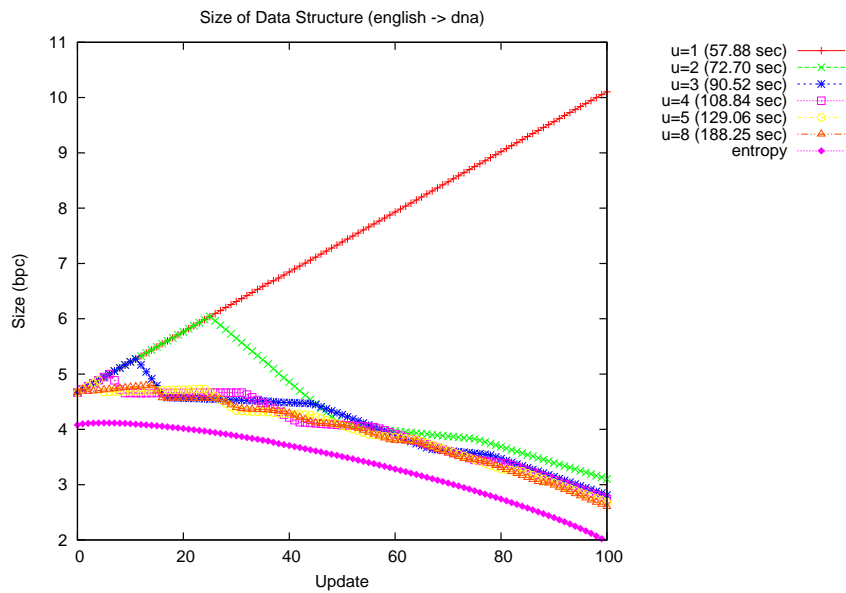}
\caption{Results for Experiment~1.
The $x$-axis represents the ratio of overwritten text;
$0$~is the initial situation where the entire text is from the English
text, $50$~means that the left half of the text has been changed to the DNA
text while the right half is still from the English text, and $100$~is
when the whole text has become the DNA text.}
\label{fig:size}
\end{center}
\end{figure}

Fig.~\ref{fig:size} shows the outcome.
As expected, when the text is modified, its $1$-st order entropy (shown in
the curve `entropy') changes.
The other lines in the diagram show the data structure's change in size for
various values of the parameter~$u$.
If $u = 1$, the code never changes because we write the whole DNA text
before the end of Phase~$1$ (see Section~\ref{sec:replace} for the
explanation of ``phase'').
In other words, all the characters in the DNA text are encoded by the optimal
code for the English text, and the resulting size is bad.
If $u > 1$, the code is updated while the text is modified, and the size
converges to the entropy (plus the size of the auxiliary data structure).
We can see that as the value of~$u$ increases, the size converges towards
the entropy more quickly but the time needed to update the data structure
also increases.
We select $u = 4$ as a good trade-off between the convergence speed and the
update time.

\medskip

\noindent
\textbf{Experiment~2}:
The second experiment compared the \texttt{access} and \texttt{replace} times
for \texttt{cram} and \texttt{cramgz}.
We set the parameters of the data structures so that both have the same size.
For \texttt{cram}, we set $b = 1024$, $m = 64$, $u = 4$.
Then, the English text has size $4.67$ bpc.
To achieve the same compression ratio ($4.67$ bpc) for \texttt{cramgz}, we
had to select $b' = 1024$.

\begin{figure}[t!]
\begin{center}
  \includegraphics[scale=0.90]{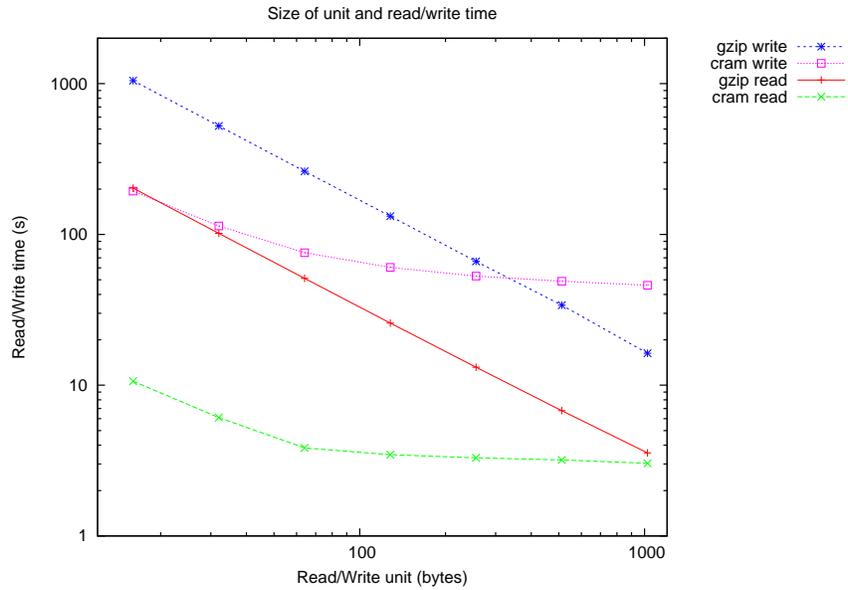}
\caption{Results for Experiment~2.
The $x$-axis shows the size of the unit for read/write, and the $y$-axis
shows the resulting time needed to read the whole text by \texttt{access}
operations or to overwrite the whole text by \texttt{replace} operations,
respectively.}
\label{fig:time}
\end{center}
\end{figure}

We performed two types of experiments:
one to evaluate \texttt{access} by measuring the time needed to read the
entire compressed English text, and one to evaluate \texttt{replace} by
measuring the time needed to overwrite the DNA text onto the compressed
English text.
We combined series of consecutive operations into single read/write unit
operations, and tried various sizes of read/write units smaller than the
block size.
Fig.~\ref{fig:time} shows the size of each read/write unit and the resulting
time for \texttt{access} and \texttt{replace}.
In \texttt{cramgz}, large blocks of length $b' = 1024$ bytes are directly
compressed by \texttt{gzip}, and reading any unit shorter than $1024$~bytes
still requires decoding the whole large block.
Therefore, \texttt{access} is not very efficient when using units shorter
than $b'$~bytes.
Similarly, writing a short unit requires decoding a large block, rewriting
a part of it, and then encoding the whole large block again.
On the other hand, in \texttt{cram}, the base is the middle block (i.e.,
$m = 64$~bytes) and therefore it is more efficient than \texttt{cramgz}.
Fig.~\ref{fig:time} shows that \texttt{cram} is faster than \texttt{cramgz}
for all read unit sizes, and faster for $16$ to $256$ bytes write unit sizes.

\end{document}